\documentclass[aps,prd,nofootinbib]{revtex4}
\usepackage{graphicx}
\usepackage{amsfonts}
\usepackage{amsmath}
\usepackage{amssymb}

\begin{document}

\title{Stability regions of glued wormholes with massless Kim-Lee backreacted spacetimes as interior}
\author{G.F. Akhtaryanova}
\email{akht\_gul@mail.ru}
\affiliation{Zel'dovich International Center for Astrophysics, Bashkir State Pedagogical University, 3A, October Revolution Street, Ufa 450077, RB, Russia}
\author{R.Kh. Karimov}
\email{karimov_ramis_92@mail.ru}
\affiliation{Zel'dovich International Center for Astrophysics, Bashkir State Pedagogical University, 3A, October Revolution Street, Ufa 450077, RB, Russia}
\author{R.N. Izmailov}
\email{izmailov.ramil@gmail.com}
\affiliation{Zel'dovich International Center for Astrophysics, Bashkir State Pedagogical University, 3A, October Revolution Street, Ufa 450077, RB, Russia}
\author{A. Bhattacharya}
\email{bamrita323@gmail.com}
\affiliation{Department of Mathematics, Kidderpore College, 2, Pitamber Sircar Lane, Kolkata 700023, WB, India}

\date{19 June 2025}

\begin{abstract}
Asymptotic zero Arnowitt-Deser-Misner (ADM) mass wormholes, such as the zero-mass traversable Ellis-Bronnikov wormhole, are of great interest for astrophysical applications such as in the galactic microlensing. However, when considered individually, they are unstable to small perturbations. On the other hand, there is a possibility that they can be stable as an interior partner of a traversable glued wormhole obtained by suitably gluing the interior to the observationally relevant massive exterior spacetimes across spherically symmetric thin shells. Although the exterior spacetime has non-zero ADM mass, massless interior partner remains massless sharing the stability of the glued wormhole. The dynamics of the thin-shell then demarcates the stability regions of the glued wormhole that we wish to study here by employing the novel concepts of thin-shell "mass" and of "external force" constraints discovered by Garcia, Lobo and Visser. We shall consider two classes, where the zero ADM mass interior are Kim-Lee wormholes glued to the exterior Schwarzschild vacuum and Reissner-Nordstr\"{o}m spacetime respectively. It turns out that the stability regions in both cases are almost similar although the two interior Kim-Lee spacetimes are physically very different, one scalar charged and the other electrically charged. The conditions under which the stability of glued wormholes could be achieved are analyzed in detail.
\end{abstract}

\maketitle


\section{Introduction}
\label{sec1}
Wormholes are geometrical handles in spacetime that connect two universes or two distant regions of spacetime and are solutions to Einstein's general relativity, including other theories of gravity and have not been disproven by experimental evidence. The concept dates back to 1918 with Flamm's paraboloid \cite{Rindler:1979}, an early geometric predecessor. Later, in 1935, Einstein and Rosen \cite{Einstein:1935} proposed a particle model representing a massless electric charge as a "bridge". However, it wasn't until 1988 that Morris and Thorne \cite{Morris:1988a} rigorously formalized the theory of wormholes in their groundbreaking work. Initially introduced as a pedagogical tool to illustrate general relativity and the possibility of interstellar travel, wormholes have since become a significant topic in modern physics. Researchers quickly recognized their broader implications, extending from quantum phenomena to classical general relativity \cite{Morris:1988b, Visser:1995}. Work in the direction of wormhole lensing has been initiated by Cramer et al. \cite{Cramer:1995}. Important studies regarding lensing of traversable wormholes \cite{Cramer:1995, Nakajima:2012, Kleihaus:2014, Izmailov:2020a, Nandi:2006}. The study \cite{Chakraborty:2022} examines the validity of two gravitational entropy proposals in traversable wormholes, finding that their applicability and consistency depend on the chosen definition, with the Clifton-Ellis-Tavakol proposal limited to Petrov type D and N spacetimes, while the geometric method applies more broadly but lacks thermodynamic relevance. In \cite{Bronnikov:2023} authors studied the properties of evolving wormholes in a closed Friedmann dust-filled universe, described by a specific branch of the Lematre--Tolman--Bondi solution and its generalization, showing their short throat lifetime, reduced matter density near the boundary, and traversability via numerical analysis of geodesics.

The recent discovery that thin-shell wormholes have the ability to mimic recently observed gravitational ring-down post-merger waves is believed to be characteristic exclusively of the black hole horizon \cite{Cardoso:2016, Cardoso:2017, Konoplya:2016, Nandi:2017}. The unified approach for analyzing idealized compact objects, such as wormholes and horizonless stars, modeled using an inner boundary wall was proposed in \cite{Sebastiani:2017}. Classically, this setup produces distinct post-ringdown echoes that help distinguish exotic objects from black holes, while quantum analysis reveals that static wormholes exhibit no radiation. Another important question is the stability of wormholes \cite{Nandi:2016} that could still be achieved using thin-shell wormholes obtained by gluing the wormholes with the some exterior spacetimes. Therefore, the study of stability of thin-shell wormholes is of paramount importance.

The first work on linearized dynamical stability analysis of the time-dependent throat of a thin shell with a prescribed equation of state was carried out by Visser \cite{Visser:1989}. The issue of stability mainly addresses the bounded motion of the wormhole throat. This dynamical analysis was generalized to linearized radial perturbations around some assumed static solutions of the Einstein field equations without the need to specify an equation of state \cite{Garcia:2012, Poisson:1995}. Possible scenarios for the evolution of a thin spherically symmetric self-gravitating phantom shell around the Schwarzschild black hole was studied in \cite{Berezin:2005}. Linear stability of asymmetric wormhole connecting Schwarzschild-Rindler spacetime to Schwarzschild-Rindler-de Sitter space-time is analyzed using this formalism in \cite{Eid:2024}. Sharif and Javed inspected the stability of thin-shell wormholes constructed by Bardeen black hole \cite{Sharif:2016} as well as Bardeen-AdS black hole \cite{Sharif:2019} by using different choices of variable EoS. The same authors \cite{Sharif:2021} compared the stable structures of thin-shell (internal Minkowski metric and the external Reissner-Nordstr\"{o}m black hole) and charged thin-shell (inner and outer Reissner-Nordstr\"{o}m black hole) wormholes using generalized barotropic, generalized phantom-like equations of state. Different types of thin-shell wormholes in $f(R)$-gravity and their stability were studied in \cite{Godani:2023,Eid:2020a,Eiroa:2016,Mazharimousavi:2020}. Stability of reflection-asymmetric thin-shell wormholes were studied in \cite{Tsukamoto:2021,Akhtaryanova:2022}. The traversability of thin-shell wormholes were studied in \cite{Celis:2021,Celis:2022,Zhang:2023}. New class of thin-shell wormholes from black holes in Ho\v{r}ava-Lifshitz gravity by employing the asymptotically flat Kehagias--Sfetsos solution with various values of the coupling constant $\omega $ and the mass $M$ were constructed and studied in \cite{Rahaman:2011}. In \cite{Banerjee:2018} authors studied the stability of higher dimensional thin-shell wormholes with a cosmological constant surrounded by quintessence. There $d$-dimensional thin-shell wormhole could have three different throat geometries: spherical, planar and hyperbolic. Intra-galactic thin shell wormhole joining two copies of identical galactic space times described by the Mannheim-Kazanas-de Sitter solution in conformal gravity and its stability under spherical perturbations studied in \cite{Bochicchio:2013}. Linear stability of thin shell wormholes joining Schwarzschild black hole with different solutions were analyzed in \cite{Khaybullina:2014,Lukmanova:2016,Akhtaryanova:2021,Karimov:2017}. Thin-shell wormholes in teleparallel-Rastall gravity were studied in \cite{Nazavari:2023}.

The present paper deals with the linear stability of thin-shell wormholes that had been studied by many researchers. The motivation for adding new ones is that the backreacted wormholes provide a new genre of objects not considered heretofore as possible interior metrics. For example, in \cite{Horowitz:2019}  authors argue that traversable wormholes can be nucleated via quantum gravity processes by creating black holes with identified horizons, which remain stationary--avoiding issues like Unruh radiation and enabling traversability through quantum field back-reaction. Intriguingly, not only is the background but also the final backreacted metric are both massless - something like preserving a zero mass gauge between the two spacetimes. Note that the zero Arnowitt -- Deser -- Misner (ADM) mass background wormhole [our Eq.(8) with $\alpha =0$, see below] is known to be linearly stable under spherically symmetric and axial perturbations in general relativity. However, the stress tensor threading the same massless wormhole consists of a perfect fluid with negative density and a source-free radial electric or magnetic field \cite{Novikov:2012}. This peculiar source distribution does provide an excellent first example threading a stable massless wormhole. On the other hand, our motivation was to consider wormholes that are more commonly known as the massless unstable Ellis-Bronnikov wormhole threaded by phantom scalar field and find the regions of stability of these wormholes via thin shell construction. The Ellis-Bronnikov phantom wormholes are massless variant of the Einstein frame version of the celebrated Brans-Dicke theory \cite{Izmailov:2020b, Khaybullina:2013}. Also wormholes with Ellis geometry have been successfully constructed in \cite{Das:2005} using tachyon matter.

Employing Garcia, Lobo and Visser (GLV) formalism \cite{Garcia:2012} the stability regions of the linearly perturbed spherical motion of the thin-shell moving in the bulk spacetime will be investigated. Particularly in this article, we wish to address the stability of the thin-shell in two models: (1) One obtained by gluing scalar field massless Kim-Lee (hereinafter KL) spacetime \cite{Kim:2001} to the vacuum Schwarzschild black hole and (2) the other obtained by gluing electrically charged massless KL spacetime \cite{Kim:2001} to the Reissner-Nordstr\"{o}m (hereinafter RN) black hole. Both glued wormholes are twice asymptotically flat with gluing being done at some suitable "standard" coordinate radius avoiding the throats and horizons. The asymptotic ADM masses in the interior will be the zero mass KL spacetimes and on the exterior will be the asymptotic Schwarzschild and RN masses respectively. We want to clarify that, although the exterior spacetime has a non-zero mass thereby producing jumps in extrinsic curvatures across the thin shell, the massless property of the interior spacetime is preserved since its asymptotic ADM is still zero. \textit{In this sense, we say that the zero-mass wormhole can be made stable, not as an individual entity but as a partner in the glued configuration.}

The paper is organized as follows: For easy references, we briefly describe in Sec.2 the zero mass KL wormhole with scalar charge. In Sec.3, we briefly describe the zero mass KL wormhole with electric charge. GLV formalism is described in Sec.4. Secs.5,6 analyze the stability regions of the two shells. Sec.7 concludes the work.

\section{KL wormhole with scalar charge}
\label{sec2}
Metric of backreacted scalar charged wormhole derived by KL \cite{Kim:2001, Kim:1996} is as follows: The Einstein equations for the background wormhole spacetime considered is 
\begin{equation}
G_{\mu \nu }^{(0)} = 8\pi T_{\mu \nu }^{(0)}.
\end{equation}%
The left hand side, $G_{\mu \nu }^{(0)}$, is the wormhole geometry and the right hand side, $T_{\mu \nu }^{(0)}$, is the exotic matter violating the known energy conditions. If additional matter $T_{\mu \nu}^{(1)}$ is added to the right hand side and the corresponding backreaction $G_{\mu \nu }^{(1)}$ is added to the left hand side, then Einstein equations become 
\begin{equation}
G_{\mu \nu }^{(0)}+G_{\mu \nu }^{(1)}=8\pi \left[ T_{\mu \nu }^{(0)}+T_{\mu\nu }^{(1)}\right] .
\end{equation}

Static Lorentzian wormhole with an additional minimally coupled massless scalar field stress-energy tensor that was considered by Kim-Lee \cite{Kim:2001} is given by 
\begin{eqnarray}
T_{\mu \nu }^{(1)} &=&\varphi _{;\mu }\varphi _{;\nu } - \frac{1}{2}g_{\mu \nu} g^{\rho \sigma}\varphi _{;\rho } \varphi _{;\sigma }, \\
\square \varphi  &=& 0
\end{eqnarray}%
where $\varphi$ is the minimally coupled scalar charge. The metric is chosen in the generic Morris-Thorne form as  
\begin{equation}
d\tau ^{2} = -e^{2\Phi (r)}dt^{2}+\left( 1 - \frac{b(r)}{r} \right)^{-1} dr^{2}+r^{2}\left( d\theta ^{2} + \sin ^{2}{\theta }d\phi ^{2}\right) ,
\end{equation}%
where $\Phi(r)$ and $b(r)$ are called redshift and shape functions respectively. For simplicity, it is assumed in \cite{Kim:2001} that $\Phi =0$. Then the backreacted metric due to scalar charge (denoted by the suffix "Sc") has been derived as the solution of Eq.(2), viz., 
\begin{equation}
d\tau_{\rm Sc}^{2} = -dt^{2} + \left(1 - \frac{b(r)}{r} + \frac{\alpha }{r^{2}}\right)^{-1}dr^{2} + r^{2}\left( d\theta ^{2}+\sin^{2}{\theta} d\phi^{2}\right) ,
\end{equation}%
where $\alpha $ plays the role of the scalar charge. For dimensional reasons, the specific form of function $b(r)$ proposed in \cite{Kim:2001} is 
\begin{equation}
b(r)=b_{0}^{2\beta /(2\beta +1)}r^{1/(2\beta +1)},
\end{equation}%
where $b_{0}$ is a constant and the parameter $\beta$ is the equation of state parameter of matter threading the background wormhole (5), which should be less than $-\frac{1}{2}$ so that the exponent of $r$ can be negative needed to satisfy the flareout condition.

The value $\beta = -1$ is chosen in \cite{Kim:2001} so that the backreacted KL wormhole with the scalar field $\varphi$ becomes 
\begin{equation}
d\tau_{\rm Sc}^{2} = -dt^{2} + \left(1 - \frac{b_{0}^{2} - \alpha}{r^{2}}\right)^{-1} dr^{2} + r^{2}\left( d\theta ^{2} + \sin^{2}{\theta} d\phi^{2}\right) 
\end{equation}%
with the scalar field 
\begin{equation}
\varphi_{\rm KL} = \varphi_{0} \left[1 - \cos^{-1} \left(\frac{b_{0}}{r}\right) \right],
\end{equation}
where the free parameter $b_{0}$ has the dimension of length. The throat of the wormhole appears at 
\begin{equation}
r_{\rm th} = \sqrt{b_{0}^{2} - \alpha},
\end{equation}%
which for the reality of the throat radius requires that $b_{0}^{2} > \alpha$ for a physically meaningful wormhole. In Ref.~\cite{Akhtaryanova:2024} it was shown that the wormhole (3) is massless, i.e. the asymptotic ADM mass $M_{\rm ADM} = 0$ that can be reduced to the massless Ellis-Bronnikov wormhole form with the replacement $m^{2} = b_{0}^{2} - \alpha$. To obtain commonly used well known form, one can use the coordinate transformation of $r^{2} = \ell^{2} + m^{2}$ \cite{Lukmanova:2016b}.

\section{KL wormhole with electric charge}
\label{sec3}
The additional stress-energy tensor for electromagnetic field is given by \cite{Kim:2001}
\begin{equation}
T_{\mu \nu}^{(1)} = \frac{1}{4\pi }\left( F_{\mu \lambda} F_{\nu }^{\lambda}\right) - \frac{1}{4} g_{\mu \nu } F_{\lambda \sigma} F^{\lambda \sigma }
\end{equation}%
and the electric field for the charge $Q$ is 
\begin{equation}
E=\frac{Q}{r^{2}}\sqrt{g_{tt}g_{rr}}.
\end{equation}

The general form of the backreacted wormhole metric due to electric charge (denoted by the suffix "El") is given by \cite{Kim:2001} 
\begin{equation}
d\tau _{\rm El}^{2} = -\left(1 + \frac{Q^{2}}{r^{2}} \right) dt^{2}+\left(1 - \frac{b(r)}{r} + \frac{Q^{2}}{r^{2}} \right)^{-1} dr^{2} + r^{2}\left( d\theta^{2}+\sin ^{2}\theta d\phi^{2}\right),
\end{equation}%
where the redshift function $\Phi =\frac{1}{2}\ln \left( 1+\frac{Q^{2}}{r^{2}}\right) $, $Q$ is the electric charge and $b(r)$ is given, as before, by 
\begin{equation}
b(r)=b_{0}^{2\beta /(2\beta +1)}r^{1/(2\beta +1)},
\end{equation}%
where $b_{0}$ is a constant. Once again, we will only consider the case $\beta =-1$ studied in \cite{Kim:2001}, which gives $b(r) = b_{0}^{2}/r$. So, the backreacted KL wormhole with electric charge can be rewritten as \cite{Kim:2001}
\begin{equation}
d\tau_{\rm El}^{2} = -\left(1 + \frac{Q^{2}}{r^{2}}%
\right) dt^{2} + \left( 1 - \frac{b_{0}^{2}}{r^{2}} + \frac{Q^{2}}{r^{2}} \right)^{-1}dr^{2}+r^{2}\left( d\theta^{2} + \sin^{2}\theta d\phi ^{2}\right) .
\end{equation}%
The throat of the wormhole is given by 
\begin{equation}
r_{\rm th} = \sqrt{b_{0}^{2}-Q^{2}}
\end{equation}%
and the quasi-local Misner-Sharp (MS) mass as $M_{\rm MS} = r_{\rm th}/2$ \cite{Misner:1964}. In order for the throat radius to exist, it is necessary to have $Q^{2}<b_{0}^{2}$. When $Q=0$, the
spacetime belongs to massless Ellis-Bronnikov wormhole and when $b_{0}=0$, it becomes the RN spacetime with zero ADM mass. It is consistent with $M_{\rm MS}\rightarrow 0$ as $r\rightarrow \infty$ as was shown in Ref.~\cite{Akhtaryanova:2024}.

\section{Thin-shell wormhole formalism of Garcia, Lobo and Visser (GLV)}
\label{sec4}
We shall only cite their main results. For details the reader is asked to read the original paper \cite{Garcia:2012}. The spacetimes on two $\pm $ sides of the thin-shell are given by
\begin{equation}
d\tau ^{2} = -e^{2\Phi _{\pm }(r_{\pm })}\left[1 - \frac{b_{\pm }(r_{\pm })}{r_{\pm }} \right] dt^{2}+\left[ 1-\frac{b_{\pm }(r_{\pm })}{r_{\pm }}\right]^{-1}dr_{\pm }^{2} + r_{\pm }^{2}d\Omega _{\pm }^{2}.
\end{equation}%
Further any two \textit{arbitrary} spherically symmetric spacetimes to be glued together by cut and paste procedure. Thus, for the static and spherically symmetric spacetime, the single manifold $\mathcal{M}$ is obtained by gluing two bulk spherically symmetric spacetimes $\mathcal{M}_{+}$ and $\mathcal{M}_{-}$ at a timelike junction surface $\sum $, i.e., at $f(r,\tau )=r-a(\tau )=0$. The surface stress-energy tensor may be written in terms of the surface energy density $\sigma $ and the surface pressure $\mathcal{P}$ as $S_{ij}=$diag ($-\sigma ,\mathcal{P},\mathcal{P}$). The general conservation law is given by
\begin{equation}
\frac{d(\sigma A)}{d\tau } + \mathcal{P} \frac{dA}{d\tau } = \Xi A\dot{a},
\end{equation}%
where $\dot{a} = \frac{da}{d\tau}$, the shell surface area $A=4\pi a^{2}$ and there is an entirely new term 
\begin{equation}
\Xi =\frac{1}{4\pi a}\left[ \Phi _{+}^{\prime}(a) \sqrt{1 - \frac{b_{+}(a)}{a} + \dot{a}^{2}}+\Phi _{-}^{\prime}(a)\sqrt{1 - \frac{b_{-}(a)}{a} + \dot{a}^{2}}\right].
\end{equation}
The first term in Eq.(18) represents the variation of the internal energy of the shell, the second term is the work done by the internal force of the shell. The right hand side is the net discontinuity in the conservation law of the surface stresses of the bulk momentum flux and is physically interpreted as the work done by external forces on the thin shell. In short it is the "external force" term occurring due to $\Phi_{\pm} \neq 0$. When $\Phi_{\pm}^{\prime }(a) = 0$, we have $\Xi = 0$, and then one recovers the familiar conservation law on the shell.

Assuming integrability of $\sigma$, which allows $\sigma =\sigma (a)$, it is possible to define the mass of the thin shell of exotic matter residing on wormhole throat as \cite{Poisson:1995}
\begin{equation}
m_{s}(a)=4\pi \sigma (a)a^{2}.
\end{equation}%
GLV derived the workable master inequalities about stability around a static radius $a_{0}$ after long calculations, which are the constraint from the "mass" 
\begin{eqnarray}
m_{s}^{\prime \prime }\left( a_{0}\right) &\geq &\frac{1}{4a_{0}^{3}}\left\{ 
\frac{\left[ b_{+}(a_{0})-a_{0}b_{+}^{\prime }(a_{0})\right] ^{2}}{\left[
1-b_{+}(a_{0})/a_{0}\right] ^{3/2}}+\frac{\left[ b_{-}(a_{0})-a_{0}b_{-}^{%
\prime }(a_{0})\right] ^{2}}{\left[ 1-b_{-}(a_{0})/a_{0}\right] ^{3/2}}%
\right\}  \nonumber \\
&&+\frac{1}{2}\left\{ \frac{b_{+}^{\prime \prime }(a_{0})}{\sqrt{%
1-b_{+}(a_{0})/a_{0}}}+\frac{b_{-}^{\prime \prime }(a_{0})}{\sqrt{%
1-b_{-}(a_{0})/a_{0}}}\right\} ,
\end{eqnarray}%
and when $\Phi _{\pm }^{\prime }(a_{0})\leq 0$, the constraint from the "external force" 
\begin{eqnarray}
\left. \left[ 4\pi a\Xi (a)\right] ^{\prime \prime }\right\vert _{a_{0}} &\geq &\left. \left\{ \Phi _{+}^{^{\prime \prime \prime }}(a)\sqrt{%
1-b_{+}(a)/a}+\Phi _{-}^{^{\prime \prime \prime }}(a)\sqrt{1-b_{-}(a)/a}\right\} \right\vert _{a_{0}}  \nonumber \\
&&-\left. \left\{ \Phi _{+}^{^{\prime \prime }}(a)\frac{\left[ b_{+}(a)/a%
\right] ^{\prime }}{\sqrt{1-b_{+}(a)/a}}+\Phi _{-}^{^{\prime \prime }}(a)%
\frac{\left[ b_{-}(a)/a\right] ^{\prime }}{\sqrt{1-b_{-}(a)/a}}\right\}
\right\vert _{a_{0}}  \nonumber \\
&&-\frac{1}{4}\left. \left\{ \Phi _{+}^{^{\prime }}(a)\frac{\left[ \left(
b_{+}(a)/a\right) ^{\prime }\right] ^{2}}{\left[ 1-b_{+}(a)/a\right] ^{3/2}}%
+\Phi _{-}^{^{\prime }}(a)\frac{\left[ \left( b_{-}(a)/a\right) ^{\prime }%
\right] ^{2}}{\left[ 1-b_{-}(a)/a\right] ^{3/2}}\right\} \right\vert _{a_{0}}
\nonumber \\
&&-\frac{1}{2}\left. \left\{ \Phi _{+}^{^{\prime }}(a)\frac{\left[ b_{+}(a)/a%
\right] ^{\prime \prime }}{\left[ 1-b_{+}(a)/a\right] }+\Phi _{-}^{^{\prime
}}(a)\frac{\left[ b_{-}(a)/a\right] ^{\prime \prime }}{\left[ 1-b_{-}(a)/a%
\right] }\right\} \right\vert _{a_{0}}.
\end{eqnarray}%
Similar, but not the same, force constraint appears for $\Phi _{\pm}^{\prime }(a_{0})\geq 0$ as well. We shall not quote it here as our example has $\Phi _{\pm }^{\prime }(a_{0})\leq 0$. Such a force contraint however disappears if $\Phi _{\pm }^{\prime }=0$.

The non-trivial components of the extrinsic curvature  are also given by GLV formalism
\begin{equation}
K_{\theta }^{\theta \pm } = \pm \frac{1}{a} \sqrt{1 - \frac{b_{\pm }(a)}{a} + \dot{a}^{2}},
\end{equation}
\begin{equation}
K_{\tau }^{\tau \pm} = \pm \left\{\frac{\bar{a} + \frac{b_{\pm }(a) - b_{\pm}^{\prime }(a) a}{2a^{2}}}{\sqrt{1-\frac{b_{\pm}(a)}{a} + \dot{a}^{2}}} + \Phi _{\pm }^{\prime }(a) \sqrt{1 - \frac{b_{\pm}(a)}{a} + \dot{a}^{2}}\right\},
\end{equation}
where the prime now denotes a derivative with respect to the coordinate $a$.

\section{Schwarzschild-zero mass scalar charged glued wormhole}
\label{sec5}
Further we glue Schwarzschild exterior metric with KL zero mass metric at some static radius $r=a_{0}>2M$ (Schwarzschild horizon), $\sqrt{b_{0}^{2}-\alpha }$ (throat radius of KL wormhole with scalar charge). The regions $r\leq 2M, \sqrt{b_{0}^{2}-\alpha }$ are surgically excised out of the respective spacetimes.

\subsection{Embedding diagram}
We will use the embedding diagrams to visualize the curvature of the thin shell Schwarzschild-Kim-Lee wormhole with a scalar charge. The diagram will consist of 3 parts: KL wormhole with scalar charge, Schwarzschild black hole and thin shell. Let us first consider the KL wormhole with scalar charge in three-dimensional space with a $t$ coordinate as constant in the equatorial plane, i.e. $\theta = \pi/2$. Then the line element (8) can be rewritten as
\begin{equation}
ds^2 = \left(1 - \frac{b_{0}^{2} - \alpha}{r^{2}} \right)^{-1} dr^{2} + r^{2} d\phi^{2}.
\end{equation}

Next, we construct the resulting slice in three-dimensional Euclidean space, which is defined by the cylindrical coordinates $z$, $r$, and $\phi$ and has the form
\begin{equation}
ds^2 = dz^{2} + dr^{2} + r^{2} d\phi^{2}.
\end{equation}

Let us rewrite equation (26) so that the embedded surface is described by the function $z = z(r)$
\begin{equation}
ds^2 = \left[1 + \left(\frac{dz}{dr}\right)^{2} \right] dr^{2} + r^{2} d\phi^{2}.
\end{equation}

Comparing equations (25) and (27), we obtain
\begin{equation}
\frac{dz^{-}}{dr} = - \sqrt{\frac{b_{0}^{2} - \alpha}{r^{2} - b_{0}^{2} + \alpha}}.
\end{equation}
Here the negative root was chosen because we consider the KL wormhole with scalar charge space as an interior. After integration we get the embedded surface function as
\begin{equation}
z^{-} (r) = - b_{0} \sqrt{1 - \frac{\alpha}{b_{0}^{2}}} \ln{\left[\frac{r}{b_{0}} + \sqrt{\frac{r^{2}}{b_{0}^{2}} - 1 + \frac{\alpha}{b_{0}^{2}}} \right]},
\end{equation}
where the constant of integration was chosen so that the expression under the logarithm is dimensionless.

For Schwarzchild black hole case the embedded surface function is given by \cite{Morris:1988a}
\begin{equation}
z^{+} (r) = 2 \sqrt{2M(r - 2M)}.
\end{equation}

To obtain embedding diagrams, it is necessary to rotate the embedding surface functions $z^{\pm} (r)$ along the vertical $z$ axis. Fig. 1 shows the embedding diagrams of the thin-shell wormhole obtained by gluing Schwarzschild exterior and KL with scalar charge interior at radius $r = a_{0} = 1.1 b_{0}$ for $\alpha = 0$, $0.5b_{0}^{2}$ and $0.9b_{0}^{2}$. To construct the embedding diagram, the relation $b_{0} = 2M$ was used. From the figure it can be seen that increasing $\alpha$ leads to a stronger curvature when moving from one space to another.

\begin{figure}
\centerline{\includegraphics[width=7.0in]{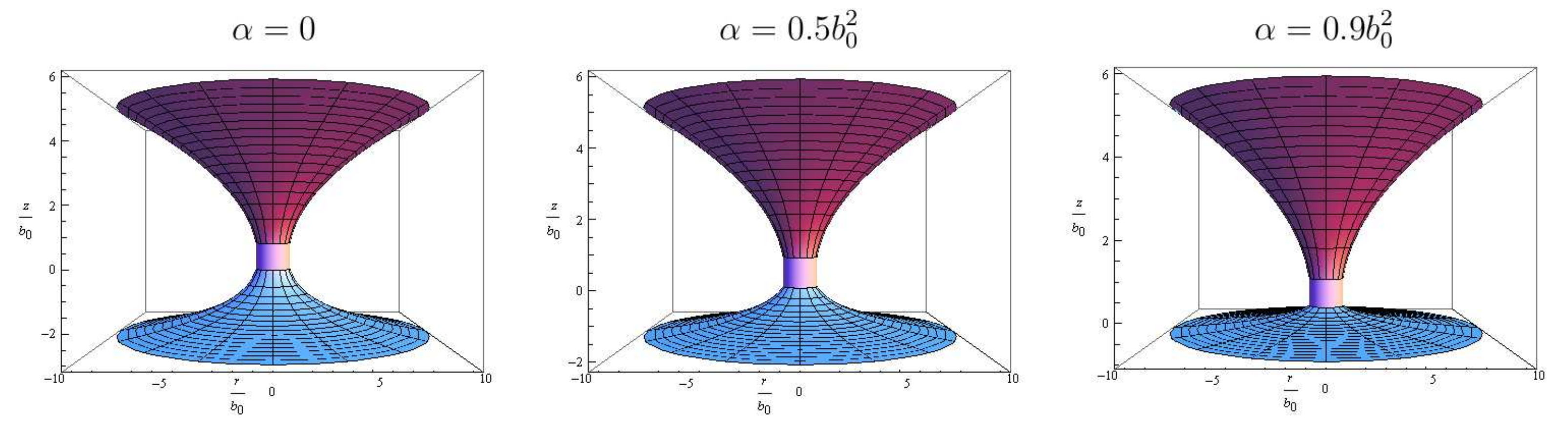}}
\caption{Embedding diagram of the thin-shell wormhole obtained by gluing Schwarzschild exterior and KL with scalar charge interior at radius $a_{0} = 1.1 b_{0}$ for different values of $\alpha$. }
\end{figure}

\subsection{Stability analisys}
Casting the Schwarzschild metric in the form (17), we get 
\begin{equation}
b_{+}=2M, \quad \Phi _{+}=0
\end{equation}%
and similarly casting KL wormhole metric (8) in the form of generic form of metric (17), we get%
\begin{equation}
b_{-}=\frac{b_{0}^{2}-\alpha }{r},\quad \Phi _{-}=-\frac{1}{2}\ln \left[ 1-%
\frac{b_{0}^{2}-\alpha }{r^{2}}\right] .
\end{equation}

Now, it is necessary to find out the sign of $\Phi _{-}^{\prime }$. To do this, by differentiating $\Phi _{-}$, we obtain 
\begin{equation}
\Phi _{-}^{\prime }=-\frac{b_{0}^{2}-\alpha }{r^{3}}\left(1 - \frac{%
b_{0}^{2}-\alpha }{r^{2}}\right) ^{-1}.
\end{equation}%
Because $b_{0}^{2}>\alpha$ then for satisfying flare out condition and using condition $\frac{b_{0}^{2}-\alpha }{r^{2}}<1$ as required by signature of metric we have 
\begin{equation}
\Phi _{-}=-\frac{b_{0}^{2}-\alpha }{r\left( r^{2}-b_{0}^{2}+\alpha \right) }%
\leq 0.
\end{equation}%
Thus we have $\Phi _{\pm }^{\prime }(a_{0})\leq 0$ for $a_{0}>m$, and there will occur the effect of "external force" influencing the thin shell motion.

The nontrivial components of the extrinsic curvature for thin shell wormhole can be obtained as
\begin{equation}
K^{\theta}_{\; \theta} = \left\{
 \begin{array}{l}
 \frac{1}{a} \sqrt{1 - \frac{2M}{a} + \dot{a}^2}, \\
 -\frac{1}{a} \sqrt{1 - \frac{b_{0}^{2} - \alpha}{a^{2}} + \dot{a}^2},
 \end{array} \right.
\end{equation}%

\begin{equation}
 K^{\tau}_{\; \tau} = \left\{
 \begin{array}{l}
 \left(\ddot{a} + \frac{M}{a^{2}}\right) \left(1 - \frac{2M}{a} + \dot{a}^2\right)^{-\frac{1}{2}}, 
 \\
 - \left(\ddot{a} + \frac{b_{0}^{2} - \alpha}{a^{3}}\right) \left(1 - \frac{b_{0}^{2} - \alpha}{a^{2}} + \dot{a}^2\right)^{-\frac{1}{2}} + \frac{b_{0}^{2} - \alpha}{a \left(a^2 - b_{0}^{2} + \alpha\right)} \sqrt{1 - \frac{b_{0}^{2} - \alpha}{a^{2}} + \dot{a}^2}.
 \end{array} \right.
\end{equation}%

Since we are considering gluing along a static radius of $a=a_{0}$, then
\begin{equation}
 K^{\theta}_{\; \theta} = \left\{
 \begin{array}{l}
 \frac{1}{a_{0}} \sqrt{1 - \frac{2M}{a_{0}}},  \\
 -\frac{1}{a_{0}} \sqrt{1 - \frac{b_{0}^{2} - \alpha}{a_{0}^{2}}}, 
 \end{array} \right.
\end{equation}%

\begin{equation}
 K^{\tau}_{\; \tau} = \left\{
 \begin{array}{l}
 \frac{M}{a_{0}^{2}} \left(1 - \frac{2M}{a_{0}}\right)^{-\frac{1}{2}}, \\
 0. 
 \end{array} \right.
\end{equation}%

Figure 2 shows the nontrivial components of the extrinsic curvature of thin-shell Schwarzschild - zero mass Kim-Lee glued wormhole with scalar charge wormhole at static radius $a=a_{0}$ for different values of $\alpha$.

The surface energy density $\sigma$, and the surface pressure $\rho$ on the junction surface can be obtained as
\begin{equation}
\sigma = - \frac{1}{4\pi a}\left[ \sqrt{1 - \frac{2M}{a} + \dot{a}^2} + \sqrt{1 - \frac{b_{0}^{2} - \alpha}{a^2} + \dot{a}^2} \right],
\end{equation} 

\begin{eqnarray}
\rho &=& - \frac{1}{8\pi a}\left[ \frac{1 + \dot{a}^{2} +a\ddot{a} - M/a}{\sqrt{1 - 2M/a + \dot{a}^{2}}} + \frac{1 + \dot{a}^{2} +a\ddot{a} - \left(b_{0}^{2} - \alpha \right)/a^{2}}{\sqrt{1 - \left(b_{0}^{2} - \alpha \right)/a^{2} + \dot{a}^{2}}} \right. \nonumber \\
&& \left.- \frac{b_{0}^{2} - \alpha}{a^{2} - b_{0}^{2} + \alpha} \sqrt{1 - \frac{b_{0}^{2} - \alpha}{a^{2}} + \dot{a}^{2}} \right],
\end{eqnarray}

Since $a=a_{0}$ is the static radius we have
\begin{equation}
\sigma = - \frac{1}{4\pi a}\left[ \sqrt{1 - \frac{2M}{a}} + \sqrt{1 - \frac{b_{0}^{2} - \alpha}{a^2}} \right],
\end{equation} 

\begin{equation}
\rho = - \frac{1}{8\pi a}\left[ \frac{1 - M/a}{\sqrt{1 - 2M/a}} + \frac{1 - \left(b_{0}^{2} - \alpha \right)/a^{2}}{\sqrt{1 - \left(b_{0}^{2} - \alpha \right)/a^{2}}} \right.  \left.- \frac{b_{0}^{2} - \alpha}{a^{2} - b_{0}^{2} + \alpha} \sqrt{1 - \frac{b_{0}^{2} - \alpha}{a^{2}}} \right],
\end{equation}

Fig.3 shows the surface energy density $\sigma$ and the surface pressure $\rho$ on the junction surface of thin-shell Schwarzschild - zero mass Kim-Lee glued wormhole with scalar charge at static radius $a=a_{0}$ for $\alpha = 0$, $0.5b_{0}^{2}$ and $0.9b_{0}^{2}$.

\begin{figure}
\centerline{\includegraphics[width=7.0in]{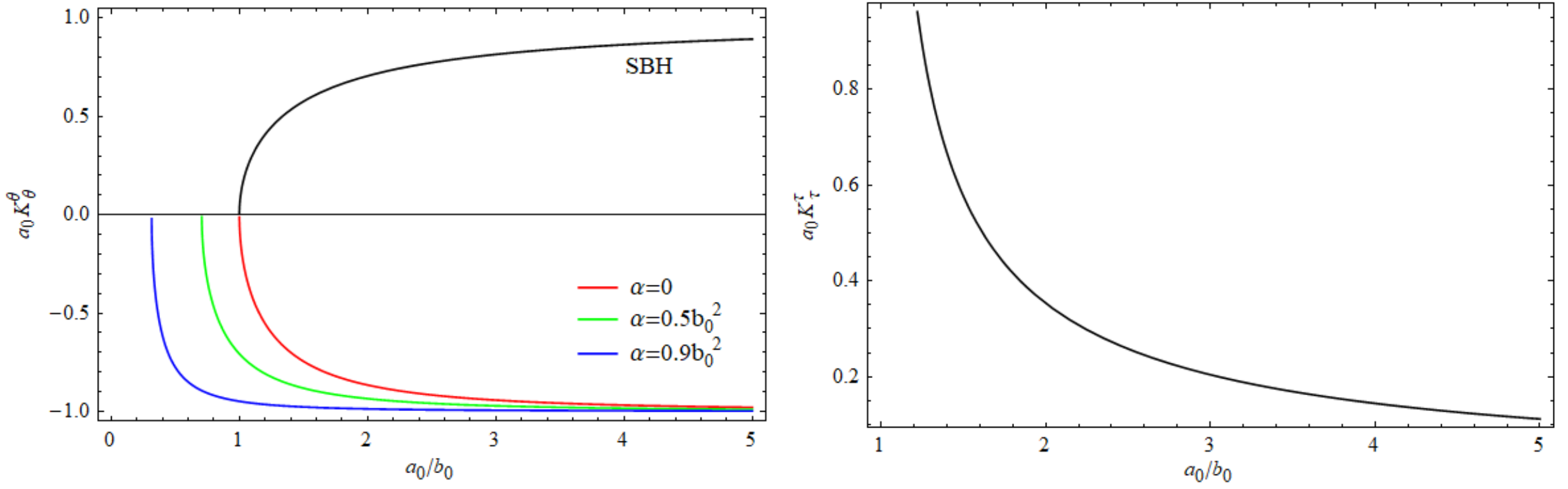}}
\caption{The nontrivial components of the extrinsic curvature of thin-shell Schwarzschild - zero mass Kim-Lee wormhole with scalar charge wormhole at static radius $a=a_{0}$ for different values of $\alpha$.}
\end{figure}

\begin{figure}
\centerline{\includegraphics[width=7.0in]{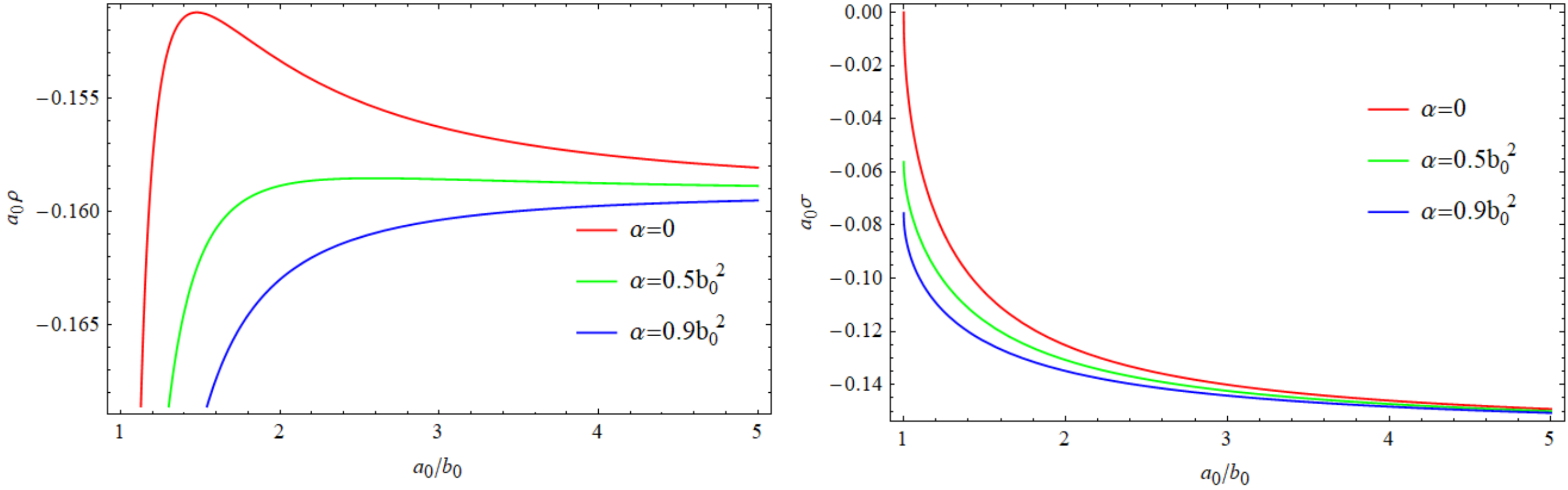}}
\caption{The surface energy density $\sigma$ and the surface pressure $\rho$ on the junction surface of thin-shell Schwarzschild - zero mass Kim-Lee wormhole with scalar charge wormhole for different values of $\alpha$.}
\end{figure}

Using the condition (34) and substituting (31), (32) into (21), (22) we obtain constraints from the "mass" and "external force" for thin-shell of Schwarzschild-zero mass Kim-Lee wormhole with scalar charge wormhole under the linear spherical perturbations 
\begin{equation}
\left. \left[ m_{s}^{\prime \prime }\left( a_{0}\right) \right] ^{\prime
\prime }\right\vert _{a_{0}}\geq \frac{(b_{0}^{2}-\alpha
)(3a_{0}^{2}-2b_{0}^{2}+2\alpha )}{a_{0}^{3}(a_{0}^{2}-b_{0}^{2}+\alpha
)^{3/2}}+\frac{M(2a_{0}-3M)}{a_{0}^{5/2}(a_{0}-2M)^{3/2}},
\end{equation}%
\begin{equation}
\left. \left[ 4\pi a\Xi (a)\right] ^{\prime \prime }\right\vert _{a_{0}}\geq 
- \frac{3(b_{0}^{2}-\alpha )\left\{
4a_{0}^{4}-5a_{0}^{2}b_{0}^{2}+2b_{0}^{4}+(5a_{0}^{2}-4b_{0}^{2})\alpha
+2\alpha ^{2}\right\} }{a_{0}^{4}\left( a_{0}^{2}-b_{0}^{2}+\alpha \right)
^{5/2}}.
\end{equation}

Using the transformation $x = \frac{M}{a_{0}}$, $y = \frac{b_{0}}{M}$ and $k = \frac{\alpha}{b_{0}^{2}}$ to go to dimensionless quantities we get 
\begin{equation}
\left. a_{0}^{2}\left[ m_{s}^{\prime \prime }\left( a_{0}\right)\right]
^{\prime \prime }\right\vert _{a_{0}} \geq f_{1}\left(x, y, k\right) = \frac{%
x(2 - 3x)}{(1 - 2x)^{3/2}} + \frac{(1 - k)x^{2}y^{2}\{3 - 2(1 - k)x^{2}y^{2}\}}{\{1
- (1 - k)x^{2}y^{2}\}^{3/2}},
\end{equation} 
\begin{equation}
\left. a_{0}^{3}\left[ 4\pi a\Xi (a)\right] ^{\prime\prime}\right\vert
_{a_{0}} \geq g_{1}\left(x, y, k\right) = -\frac{3 (1 - k) x^{2}
y^{2}\left\{4 + 2 (1 - k)^{2} x^{4} y^{4} - 5 (1 - k) x^{2} y^{2}\right\}}{%
\left\{1 - (1 - k) x^{2} y^{2}\right\}^{5/2}}.
\end{equation}

\begin{figure}
\centerline{\includegraphics[width=7.0in]{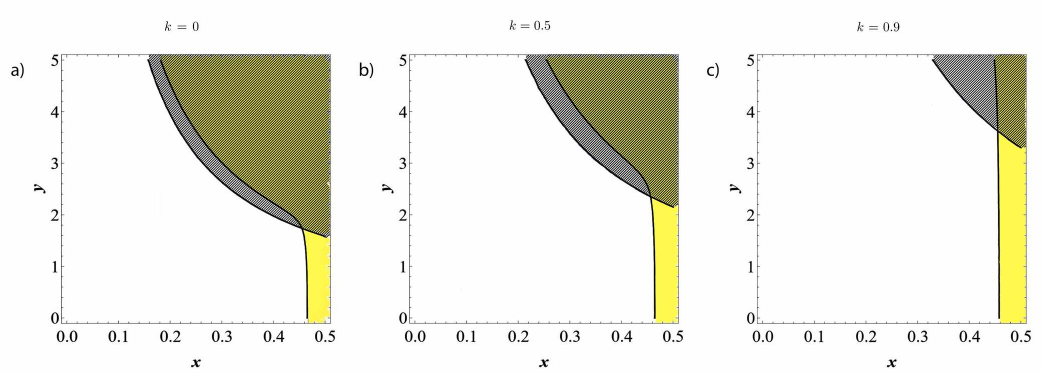}}
\caption{Stability regions of the thin-shell wormhole obtained by gluing Schwarzschild exterior and KL with scalar charge interior represented by the shaded yellow area: yellow area -- obtained from the "mass" constraint, shaded area -- obtained from the "external force" constraint.}
\end{figure}

The stability region of the thin shell Schwarzschild-Kim-Lee glued wormhole with a scalar charge is determined by conditions (19) and (20), which represent the "mass" $f_{1}$ and "external force" $g_{1}$ constraints, respectively. In Fig. 4a,b,c, the stability region obtained from the "mass" constraint is colored yellow, and the stability region obtained from the "external force" constraint is shown by the shaded region. The thin shell will be stable if both conditions are met, i.e. the stability region in Fig. 4a, b, c will be shown by the shaded yellow region. Fig. 4a shows the case when the dimensionless scalar charge is set to zero $k=0$. This special case corresponds to the Ellis-Bronnikov wormhole \cite{Ellis:1973,Bronnikov:1973}, the stability analysis of which was described in detail in Ref.~\cite{Khaybullina:2014}. Fig. 4b shows the case when the scalar charge increases to $k=0.5$. In this case, the stability region decreases by 21 percent of the maximum value (corresponds to the case $k=0$). The case $k=0.9$ is shown in Fig. 4c. In this case, the stability region is only 38 percent of the maximum value. And the stability region itself is defined for $x>0.45$ and $y>3.5$. Thus, an increase in the scalar charge leads to a decrease in the stability region of the thin shell of the Schwarzschild-Kim-Lee wormhole with a scalar charge.

\section{RN-zero mass electric charged glued wormhole}
\label{sec6}
Here we glue RN exterior metric with Kim-Lee zero mass metric at some static radius $r=a_{0}>M\pm \sqrt{M^{2}-Q_{RN}^{2}}$ (event horizon radius of RN black hole), $\sqrt{b_{0}^{2}-Q_{KL}^{2}}$ (throat radius of KL wormhole with electric charge). The regions $r\leq M\pm \sqrt{M^{2}-Q_{RN}^{2}}$ and $\sqrt{b_{0}^{2}-Q_{KL}^{2}}$ are surgically excised out of the respective spacetimes.

\subsection{Embedding diagram}
Now, we discuss embedding surface diagrams for RN-zero mass electric charged KL thin shell wormhole with $t=const.$ and $\theta = 2\pi$. Within the constraint $t=const.$ and $\theta = 2\pi$,we have the following relation from metric (15):
\begin{equation}
ds^2 = \left(1 - \frac{b_{0}^{2} - Q_{KL}^{2} }{r^{2}} \right)^{-1} dr^{2} + r^{2} d\phi^{2}
\end{equation}
We will construct two-dimensional surface of thin shell wormhole in three-dimensional Euclidean space. Comparing equations (47) and (27), we obtain 
\begin{equation}
\frac{dz}{dr} = - \sqrt{\frac{b_{0}^{2} - Q_{KL}^{2}}{r^{2} - b_{0}^{2} + Q_{KL}^{2}}}.
\end{equation}
After integration the embedded surface function of KL wormhole with electric charge given as
\begin{equation}
z (r) = - b_{0} \sqrt{1 - \frac{Q_{KL}^{2}}{b_{0}^{2}}} \ln{\left[\frac{r}{b_{0}} + \sqrt{\frac{r^{2}}{b_{0}^{2}} - 1 + \frac{Q_{KL}^{2}}{b_{0}^{2}}} \right]}.
\end{equation}
For RN solution the Embedding diagram is given in Refs. \cite{Sadegh:2024, Paranjape:2004}  and have form $z (r) = z (r, M, Q_{RN})$. Fig. 5 shows the embedding diagrams of the thin-shell wormhole obtained by gluing Schwarzschild exterior and KL with electric charge interior at radius $r = a_{0} = 1.1 b_{0}$ for different values of $Q_{KL}$ and $Q_{RN}$. To construct the embedding diagram, the relation $b_{0} = 2M$ was used.

\begin{figure}
\centerline{\includegraphics[width=7.0in]{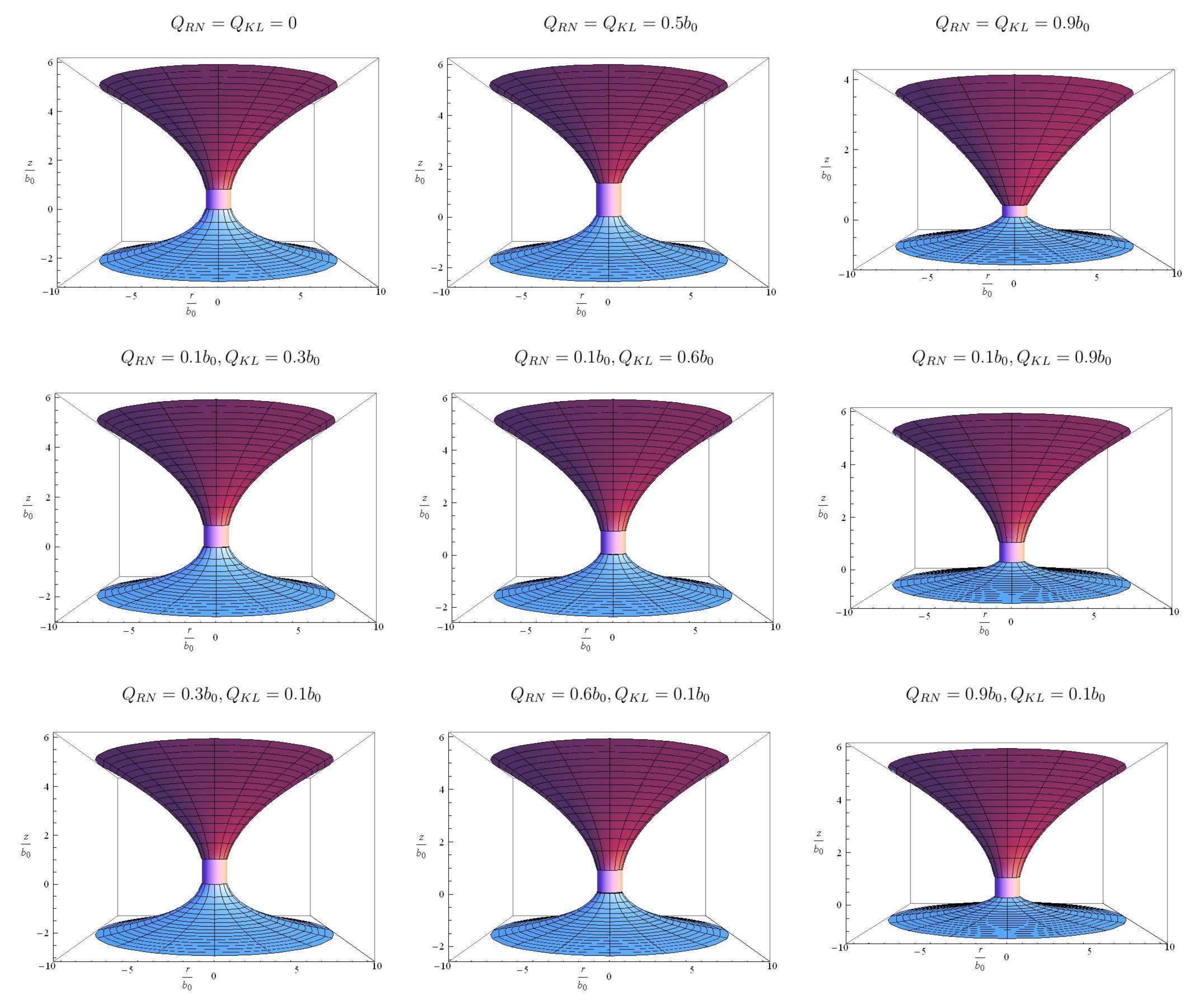}}
\caption{Embedding diagram of the thin-shell wormhole obtained by  gluing RN exterior and KL with electric charge interior at radius $a_{0} = 1.1 b_{0}$ for different values of $\alpha$.}
\end{figure}

\subsection{Stability analisys}
Casting the RN metric in the form (17), we get%
\begin{equation}
b_{+}=2M-\frac{Q_{RN}^{2}}{r},\quad \Phi _{+}=0
\end{equation}%
and similarly casting KL electrically charged wormhole metric (15) in the form of metric (17), we get%
\begin{equation}
b_{-}=\frac{b_{0}^{2}-Q_{KL}^{2}}{r},\quad \Phi _{-}=\frac{1}{2}\log \left[
\left( 1+\frac{Q_{KL}^{2}}{r^{2}}\right) \left( 1-\frac{b_{0}^{2}-Q_{KL}^{2}%
}{r^{2}}\right) ^{-1}\right] .
\end{equation}%
From (51) we can find 
\begin{equation*}
\Phi _{-}^{\prime }=-\frac{b_{0}^{2}-Q_{KL}^{2}}{\left(
r^{2}-b_{0}^{2}+Q_{KL}^{2}\right) \left( r^{2}+Q_{KL}^{2}\right) }.
\end{equation*}

The nontrivial components of the extrinsic curvature for thin shell wormhole can be obtained as 
\begin{equation}
K^{\theta}_{\; \theta} = \left\{
 \begin{array}{l}
 \frac{1}{a} \sqrt{1 - \frac{2M}{a} + \frac{Q_{RN}^{2}}{a^{2}} + \dot{a}^2},  \\
 -\frac{1}{a} \sqrt{1 - \frac{b_{0}^{2} - Q_{KL}^{2}}{a^{2}} + \dot{a}^2}, 
 \end{array} \right.
\end{equation}%

\begin{equation}
 K^{\tau}_{\; \tau} = \left\{
 \begin{array}{ll}
 \left(\ddot{a} + \frac{Ma - Q_{RN}^{2}}{a^{3}}\right) \left(1 - \frac{2M}{a} + \frac{Q_{RN}^{2}}{a^{2}} + \dot{a}^2\right)^{-\frac{1}{2}} 
 & \textrm{SBH} \textrm{,}\\
 - \left(\ddot{a} + \frac{b_{0}^{2} - Q_{KL}^{2}}{a^{3}}\right) \left(1 - \frac{b_{0}^{2} - Q_{KL}^{2}}{a^{2}} + \dot{a}^2\right)^{-\frac{1}{2}} + 
\frac{a b_{0}^{2} \sqrt{1 - (b_{0}^{2} - Q_{KL}^{2})/a^{2} + \dot{a}^2} }{\left(a^2 + Q_{KL}^{2}\right) \left(a^2 - b_{0}^{2} + Q_{KL}^{2}\right)}
 & \textrm{KLWH with scalar charge.}
 \end{array} \right.
\end{equation}%

Since $a=a_{0}$ is the static radius
\begin{equation}
 K^{\theta}_{\; \theta} = \left\{
 \begin{array}{l}
 \frac{1}{a_{0}} \sqrt{1 - \frac{2M}{a_{0}} + \frac{Q_{RN}^{2}}{a_{0}^{2}}},  \\
 -\frac{1}{a_{0}} \sqrt{1 - \frac{b_{0}^{2} - Q_{KL}^{2}}{a_{0}^{2}}}, 
 \end{array} \right.
\end{equation}%

\begin{equation}
 K^{\tau}_{\; \tau} = \left\{
 \begin{array}{l}
 \left(\frac{Ma - Q_{RN}^{2}}{a^{3}}\right) \left(1 - \frac{2M}{a} + \frac{Q_{RN}^{2}}{a^{2}}\right)^{-\frac{1}{2}},  \\
 - \frac{b_{0}^{2} - Q_{KL}^{2}}{a^{2} \sqrt{a^2 - b_{0}^{2} + Q_{KL}^{2}}} + 
\frac{b_{0}^{2}}{\left(a^2 + Q_{KL}^{2}\right) \sqrt{a^2 - b_{0}^{2} + Q_{KL}^{2}}},
 \end{array} \right.
\end{equation}%

The nontrivial components of the extrinsic curvature of thin-shell Schwarzschild - zero mass Kim-Lee glued wormhole with scalar charge wormhole at static radius $a=a_{0}$ for different values of $Q_{KL}$ and $Q_{RN}$ are shown in Fig.6.

The surface energy density $\sigma$, and the surface pressure $\rho$ on the junction surface can be obtained as
\begin{equation}
\sigma = - \frac{1}{4\pi a}\left[ \sqrt{1 - \frac{2M}{a} + \frac{Q_{RN}^{2}}{a^{2}} + \dot{a}^2} + \sqrt{1 - \frac{b_{0}^{2} - Q_{KL}^{2}}{a^2} + \dot{a}^2} \right],
\end{equation} 

\begin{eqnarray}
\rho &=& - \frac{1}{8\pi a}\left[ \frac{1 + \dot{a}^{2} +a\ddot{a} - M/a}{\sqrt{1 - 2M/a + Q_{RN}^{2}/a^{2} + \dot{a}^{2}}} + \frac{1 + \dot{a}^{2} +a\ddot{a} - \left(b_{0}^{2} - Q_{KL}^{2} \right)/a^{2}}{\sqrt{1 - \left(b_{0}^{2} - Q_{KL}^{2} \right)/a^{2} + \dot{a}^{2}}} \right. \nonumber \\
&& \left.- \frac{a^{2} b_{0}^{2}}{\left(a^{2} + Q_{KL}^{2}\right) \left(a^{2} - b_{0}^{2} + Q_{KL}^{2}\right)} \sqrt{1 - \frac{b_{0}^{2} - Q_{KL}^{2}}{a^{2}} + \dot{a}^{2}} \right],
\end{eqnarray}

Since $a=a_{0}$ is the static radius we have
\begin{equation}
\sigma = - \frac{1}{4\pi a_{0}}\left[ \sqrt{1 - \frac{2M}{a_{0}} + \frac{Q_{RN}^{2}}{a_{0}^{2}}} + \sqrt{1 - \frac{b_{0}^{2} - Q_{KL}^{2}}{a_{0}^2}} \right],
\end{equation} 

\begin{equation}
\rho = - \frac{1}{8\pi a_{0}}\left[ \frac{1 - M/a_{0}}{\sqrt{1 - 2M/a_{0} + Q_{RN}^{2}/a_{0}^{2}}} + \frac{1 - \left(b_{0}^{2} - Q_{KL}^{2} \right)/a_{0}^{2}}{\sqrt{1 - \left(b_{0}^{2} - Q_{KL}^{2} \right)/a_{0}^{2}}} - \frac{a_{0} b_{0}^{2}}{\left(a_{0}^{2} + Q_{KL}^{2}\right) \sqrt{a_{0}^{2} - b_{0}^{2} + Q_{KL}^{2}}}\right],
\end{equation}

In Fig. 7 shown the surface energy density $\sigma$ and the surface pressure $\rho$ on the junction surface of thin-shell Schwarzschild - zero mass Kim-Lee glued wormhole with electric charge wormhole at static radius $a=a_{0}$ for different values of $Q_{KL}$ and $Q_{RN}$.

\begin{figure}
\centerline{\includegraphics[width=7.0in]{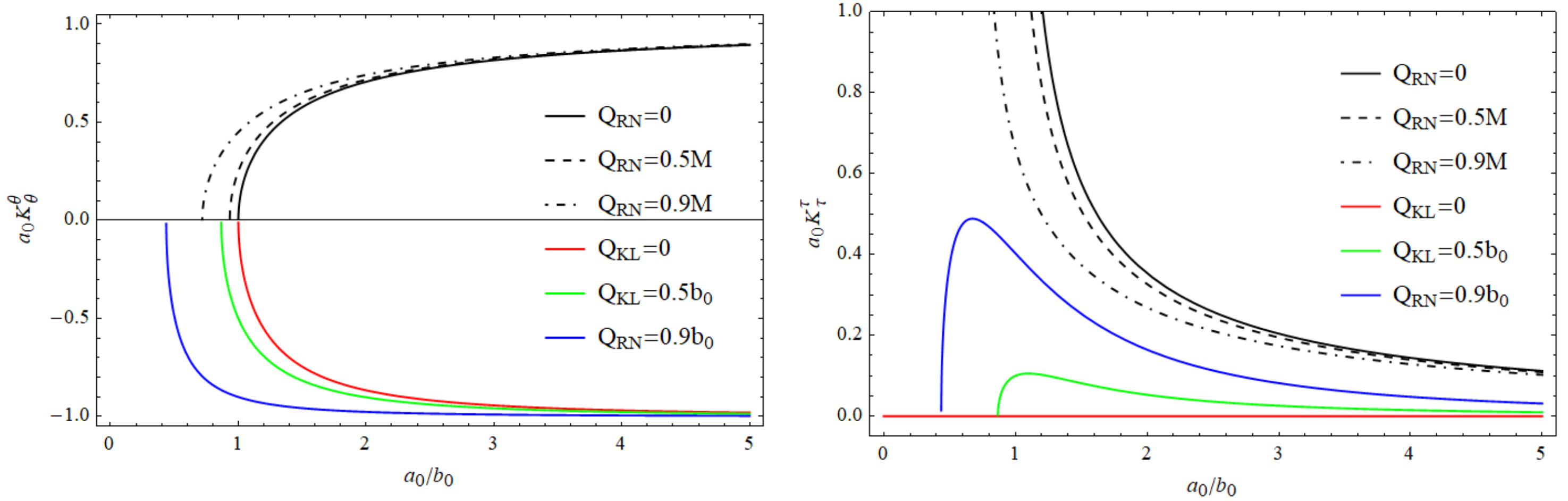}}
\caption{The nontrivial components of the extrinsic curvature of thin-shell Schwarzschild - zero mass Kim-Lee wormhole with scalar charge wormhole at static radius $a=a_{0}$ for different values of $\alpha$.}
\end{figure}

\begin{figure}
\centerline{\includegraphics[width=7.0in]{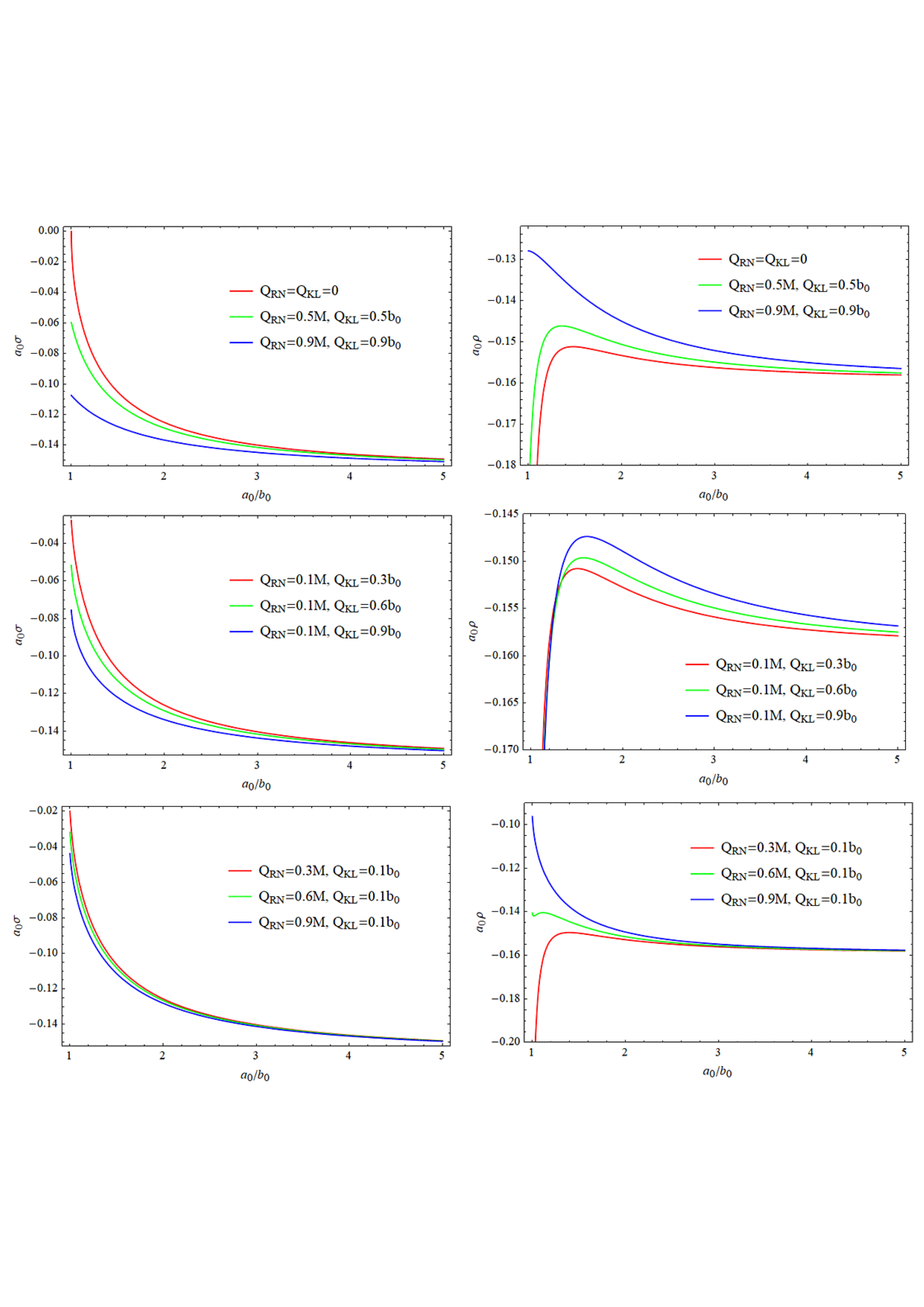}}
\caption{The surface energy density $\sigma$ and the surface pressure $\rho$ on the junction surface of thin-shell Schwarzschild - zero mass Kim-Lee wormhole with scalar charge wormhole for different values of $\alpha$.}
\end{figure}

\begin{figure}
\centerline{\includegraphics[width=7.0in]{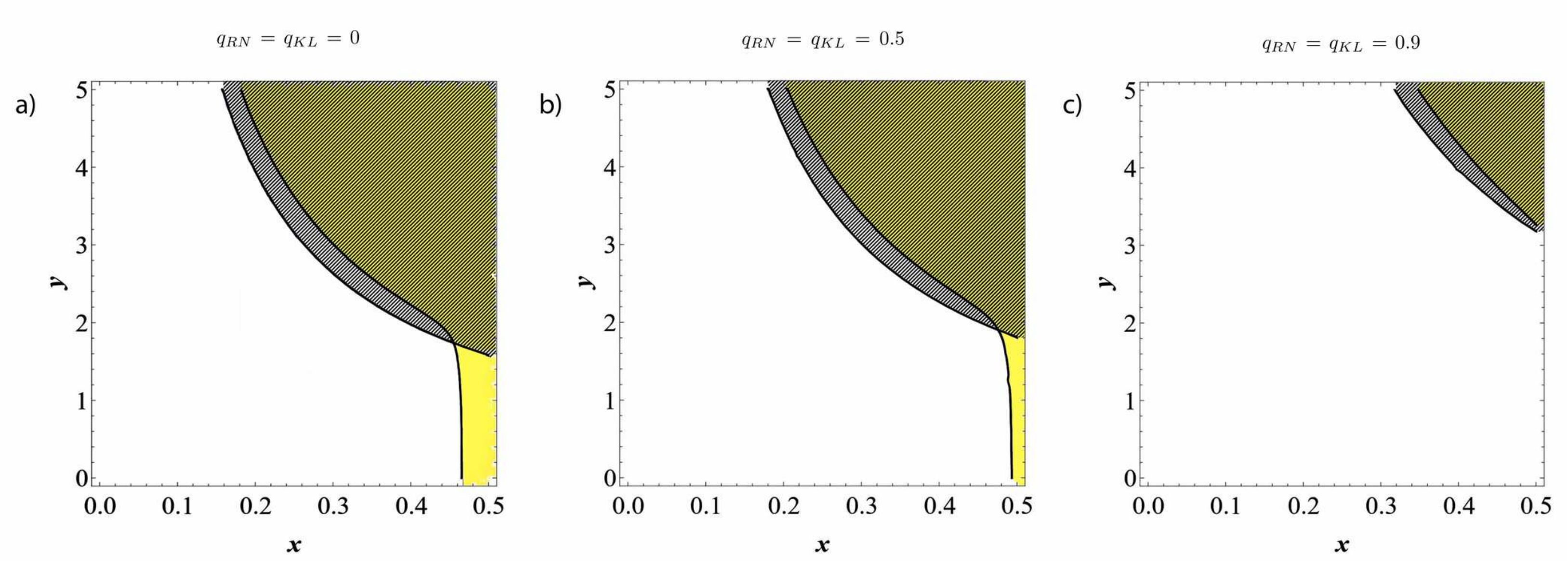}}
\caption{Stability regions of the thin-shell wormhole obtained by gluing RN exterior and KL with electric charge interior with $q_{RN}=q_{KL}$ represented by the shaded yellow area: yellow area -- obtained from the "mass" constraint, shaded area -- obtained from the "external force" constraint.}
\end{figure}

\begin{figure}
\centerline{\includegraphics[width=7.0in]{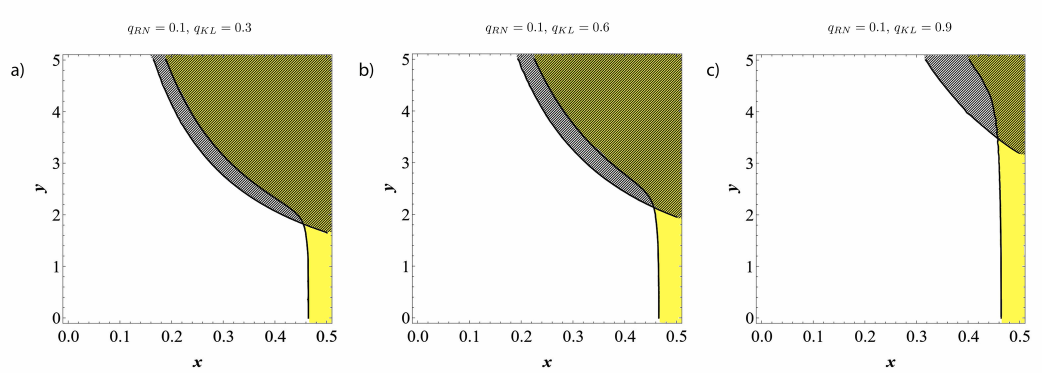}}
\caption{Stability regions of the thin-shell wormhole obtained by gluing RN exterior and KL with electric charge interior with $q_{RN}<q_{KL}$ represented by the shaded yellow area: yellow area -- obtained from the "mass" constraint, shaded area -- obtained from the "external force" constraint.
\label{fig3}}
\end{figure}

\begin{figure}
\centerline{\includegraphics[width=7.0in]{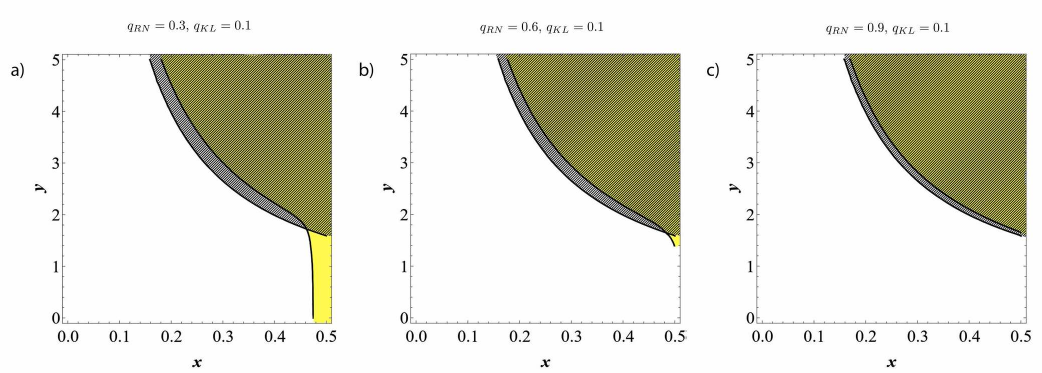}}
\caption{Stability regions of the thin-shell wormhole obtained by gluing RN exterior and KL with electric charge interior with $q_{RN}>q_{KL}$ represented by the shaded yellow area: yellow area -- obtained from the "mass" constraint, shaded area -- obtained from the "external force" constraint.
\label{fig4}}
\end{figure}

Because $r>0$ then for satisfying flare out condition and using condition $r^{2} > b_{0}^{2} - Q_{KL}^{2}$ as required by signature of metric we have 
\begin{equation}
\Phi _{-}^{\prime }\geq 0,
\end{equation}%
therefore 
\begin{eqnarray}
\left.\left[ m_{s}^{\prime \prime}\left(a_{0}\right)\right]%
^{\prime\prime}\right\vert _{a_{0}} &=& \frac{1}{r^{3}} \left[ \frac{ \left(
b_{0}^{2} - Q_{KL}^{2}\right) ^{2}}{ \left(r^{2} - b_{0}^{2} +
Q_{KL}^{2}\right) ^{3/2}} + \frac{\left(Mr - Q_{RN}^{2}\right) ^{2}}{%
\left(r^{2} - 2Mr + Q_{RN}^{2}\right) ^{3/2}} \right.  \nonumber \\
&& \left. + \frac{3(b_{0}^{2} - Q_{KL}^{2})}{\sqrt{r^{2} - b_{0}^{2} +
Q_{KL}^{2}}} + \frac{2Mr - 3Q_{RN}^{2}}{\sqrt{r^{2} - 2Mr + Q_{RN}^{2}}} %
\right] ,
\end{eqnarray}
\begin{eqnarray}
\left.\left[ 4\pi a\Xi (a)\right] ^{\prime \prime}\right\vert _{a_{0}} &=& 
\frac{ b_{0}^{2} \left[ 2b_{0}^{4} (3r^{2} - Q_{KL}^{2}) - 5
b_{0}^{2}\left(3r^{4} + 2Q_{KL}^{2}r^{2} + Q_{KL}^{4}\right)\right]}{\left(
r^{2} + Q_{KL}^{2}\right)^{3} \left(r^{2} - b_{0}^{2} + Q_{KL}^{2}\right)
^{5/2}}  \nonumber \\
&& + \frac{3 b_{0}^{2} \left(4r^{2} - Q_{KL}^{2}\right)}{\left( r^{2} +
Q_{KL}^{2}\right) \left(r^{2} - b_{0}^{2} + Q_{KL}^{2}\right) ^{5/2}}.
\end{eqnarray}

Defining 
\begin{equation}
\frac{M}{r}=x,\;\frac{b_{o}}{M}=y,\;\frac{Q_{RN}}{M}=q_{RN},\;\frac{Q_{KL}}{%
b_{0}}=q_{KL}
\end{equation}%
we find 
\begin{eqnarray}
\left. a^{2}\left[ m_{s}^{\prime \prime }\left( a_{0}\right) \right]
^{\prime \prime }\right\vert _{a_{0}} &\geq& f_{2}\left( x,y,q\right)  \nonumber
\\
&=&\frac{2x}{\sqrt{1-x\left( 2-q_{RN}^{2}x\right) }}+x^{2}\left[ \frac{%
1-q_{RN}^{2}(3-4x+2q_{RN}^{2}x^{2})}{\left\{ 1-x\left( 2-q_{RN}^{2}x\right)
\right\} ^{3/2}}\right.  \nonumber \\
&&\left. +\frac{y^{2}\left( 1-q_{KL}^{2}\right) \left\{ 3-2\left(
1-q_{KL}^{2}\right) x^{2}y^{2}\right\} }{\left\{ 1-\left(
1-q_{KL}^{2}\right) x^{2}y^{2}\right\} ^{3/2}}\right] ,
\end{eqnarray}
\begin{eqnarray}
\left. a^{3}\left[ 4\pi a\Xi (a)\right] ^{\prime \prime }\right\vert
_{a_{0}}&\geq& g_{2}\left( x,y,q\right)  \nonumber \\
&=& x^2 y^2 \left[ \frac{- 12 + 3 \left(5 - 7 q_{KL}^2\right) x^2 y^2 - 2
\left(3 - 5 q_{KL}^2 + 3 q_{KL}^4 \right) x^4 y^4}{\left( 1 + q_{KL}^2 x^2
y^2\right)^3 \left\{1 - \left(1 - q_{KL}^2\right) x^2 y^2\right\}^{5/2}}
\right.  \nonumber \\
&& \left.+ \frac{q_{KL}^2 \left(2 - 5 q_{KL}^{2} + 3 q_{KL}^{4}\right) x^6
y^6}{\left( 1 + q_{KL}^2 x^2 y^2\right)^3 \left\{1 - \left(1 -
q_{KL}^2\right) x^2 y^2\right\}^{5/2}} \right].
\end{eqnarray}

Since we are gluing two geometries with different charges, it would be interesting to consider how different charge ratios affect the stability region. Further in the paper we consider three special cases:

1) $q_{RN}=q_{KL}$, i.e. the case when the charge of the RN manifold is equal to the charge of the KL manifold;

2) $q_{RN}<q_{KL}$, i.e. the case when the charge of the RN manifold is less than the charge of the KL manifold;

3) $q_{RN}>q_{KL}$, i.e. the case when the charge of the RN manifold is greater than the charge of the KL manifold.

The stability region of the thin shell of the RN-KL wormhole with an electric charge is determined by conditions (64) and (65), which are the "mass" and "external force" constraints, respectively. Fig. 8a, b, c illustrates the case when the charge $q_{RN}$ is equal to $q_{KL}$, i.e., the case when the charge of the RN black hole is equal to the charge of the KL wormhole. The stability region obtained from the "mass" constraint is colored yellow, and the stability region obtained from the "external force" constraint condition is shown as the shaded region. Thus, the stability region is defined by two conditions and, accordingly, will be the yellow shaded region. The case when the charge $q_{RN}=q_{KL}=0$ is shown in Fig. 8a and corresponds to the Schwarzschild-Ellis-Bronnikov wormhole \cite{Ellis:1973, Bronnikov:1973}. As the charges increase to $q_{RN}=q_{KL}=0.5$, the stability region decreases by 7 percent of the maximum value (corresponds to the case $q_{RN}=q_{KL}=0$), as shown in Fig. 8b. The case $q_{RN}=q_{KL}=0.9$ is shown in Fig. 8c. In this case, the stability region decreased by 47 percent of the maximum value. Thus, when we took the same charges of the manifolds on both sides of the thin shell, increasing the charge led to a decrease in the stability region, i.e. a decrease in the admissible values of the cross-linking radius.

Fig. 9a, b, c shows the case when the charge $q_{RN}$ is equal to $q_{KL}$, i.e. the case when the charge $q_{RN}$ is less than $q_{KL}$, i.e. the case when the charge of the black hole RN is less than the charge of the wormhole KL. As in the previous cases, the stability region obtained from the "mass" constraint is colored yellow, and the stability region obtained from the "external force" constraint condition is shown as the shaded region. Therefore, the stability region is determined by two conditions and, accordingly, will be a yellow shaded region. The charge of the black hole RN is chosen to be $q_{RN}=0.1$, and the charge of the KL wormhole $q_{KL}$ will vary from $0.3$ to $0.9$. The case when the charge $q_{RN}=0.1$ and $q_{KL}=0.3$ is shown in Fig. 9a. The stability region of the thin shell is defined over the surfaces $f_{2}$ and $g_{2}$, i.e., the yellow shaded region. The stability region corresponds to the values $x>0.16$ and $y>1.7$. Increasing the charge $q_{KL}$ to $0.5$ leads to a decrease in the stability region of the thin shell by 8 percent, as shown in Fig. 9b. The case $q_{KL}=0.9$ is shown in Fig. 9c. In this case, the stability region becomes 53 percent smaller. Thus, when considering the case when the manifold RN on one side has a smaller charge compared to the other manifold KL, the stability region of the thin shell decreases.

Fig. 10a, b, c shows the case when the charge $q_{RN}$ is greater than $q_{KL} $, that is, the case when the charge of the black hole RN is greater than the charge of the wormhole KL. The charge of the black hole RN $q_{RN}$ will vary from $0.3$ to $0.9$, and the charge of the wormhole KL will be chosen equal to $q_{KL}=0.1$. The case when the charge $q_{RN}=0.3$ and $q_{KL}=0.1$ is shown in Fig. 10a. The stability region of the thin shell is defined over the surfaces $f_{2}$ and $g_{2}$. The general stability region is shown by the yellow shaded area. Increasing the charge $q_{RN}$ to $0.6$ leads to an increase in the stability region of the thin shell by 1.1 percent of the initial value (the case $q_{RN}=0.3$ and $q_{KL}=0.1$), as shown in Fig. 10b. The case $q_{RN}=0.9$ is shown in Fig. 10c. In this case, the stability region has become larger by 5.5 percent. Thus, when considering the case when one RN manifold has a higher charge compared to the other manifold, the stability region of the thin shell increases.

\section{Conclusions and summary}
\label{sec5}
The scalar charged backreacted KL wormhole (8) is formally the same as the familiar but unstable massless Ellis-Bronnikov wormhole, whose microlensing signatures i.e. the Paczy\'{n}skii light curves and the probabilistic features such as optical depth and event rate in the galactic halo have been studied in \cite{Abe:2010, Lukmanova:2016b} as possible candidates for dark halo objects. With reagard to the instability, it was shown by Novikov and Shatskiy \cite{Novikov:2012} that if the minimally coupled phantom scalar field threading the massless Ellis-Bronnikov wormhole is replaced by a radial magnetic field and an ideal phantom-fluid, then the wormhole with Ellis-Bronnikov wormhole metric could be stable. In this paper, we showed that there is yet another possibility -- stability of the KL wormhole can be achieved if it plays the role of an interior partner in a glued wormhole across a thin-shell that preserves the zero asymptotic mass of the interior. The dynamics of the thin-shell then determines the stability regions on the parametric ($x,y$) plane. This is the main idea of the paper. Specifically, we have investigated here the stability of two models of glued wormholes: the Schwarzschild-KL wormhole with a scalar charge (model 1) and the RN-KL wormhole with electric charge (model 2). The stability region of the thin-shell of the glued wormhole of model 1 is determined by two surfaces in the ($x,y$) parametric plane determined by the mass constraint and the external force constraint, while the overall stability region is determined by both the surfaces, as shown in Fig.4. The figures show that increasing the scalar charge $\alpha $ from $0$ to $0.9b_{0}^{2}$ leads to a decrease in the stability region of the wormhole.

More interesting results were obtained for the model 2, where exterior RN black hole has an electric charge $q_{RN}$ and its interior partner massless KL wormhole has an electric charge $q_{KL}$. The stability region of the thin shell is shown in Figs. 8 and 9, and the overall stability region is determined by both the surfaces, shown in Figs. 10. Three possible special cases are considered (i) $q_{RN}=q_{KL}$ (Fig. 8) and (ii) $q_{RN}<q_{KL}$ (Fig.9) and (iii) $q_{RN}>q_{KL}$ (Fig. 10). Figs. 8 show that an increase in the charges lead to a decrease in the stability region. In case (ii), when $q_{RN}$ is kept fixed, but the $q_{KL}$ is increased, the stability region is decreased. In the case (iii), when $q_{RN}$ increases, and the $q_{KL}$ is fixed, it leads to an increase in the stability region of the glued wormhole. Thus we conclude that, while massless wormholes could be individually unstable, as an interior partner of a glued wormhole, they could be stable within the parametric regions discussed above.

In summary, we state that KL wormholes could be the most intriguing examples of wormholes that are born out of cut and paste surgery and that remain massless even after the effect of backreaction. Therefore, the implication of the present work is to understand their stability features and the goal of the present paper was to throw light on staibilty aspects of these objects. We discover that only for some limited some parameter values, the wormholes are stable.as evidenced by the contour plots.

\end{document}